\documentclass[a4paper,10pt]{article}
\usepackage[utf8]{inputenc}
\usepackage{amsthm}

\usepackage{amssymb}
\usepackage{graphicx}
\usepackage{bm}
\usepackage{amsmath}
\usepackage{longtable}
\usepackage{lscape} 
\usepackage{color}
\usepackage[labelfont=bf,labelsep=period,justification=raggedright]{caption}

\usepackage{hyperref}

\usepackage[margin=1.3in]{geometry}

\title{\bf Digraphs are different:\\Why directionality matters in complex systems}
\author{Samuel Johnson
\\
\small{School of Mathematics, University of Birmingham,} 
\small{Edgbaston B15 2TT, UK;}\\
\small{The Alan Turing Institute, British Library,}
\small{96 Euston Rd, London NW1 2DB, UK.
}\\
\small{S.Johnson.4@bham.ac.uk}\\
}
\date{}

\begin{document}

      \maketitle


\begin{abstract}
\noindent
Many networks describing complex systems are directed: the interactions between elements 
are not symmetric. Recent work has shown that these networks can display properties
such as {\it trophic coherence} or {\it non-normality}, which in turn 
affect stability, percolation and other dynamical features.
I show here that these topological properties have a common origin, in that
the edges of directed networks can be aligned -- or not -- with a global direction.
And I illustrate how this can lead to rich and unexpected dynamical behaviour even 
in the simplest of models.
\end{abstract}

\section*{The importance of being directed}
\noindent
Complex systems -- be they cells, ecosystems, brains or financial markets -- 
are invariably made up of many elements interacting 
in non-trivial ways.
A simple yet powerful description of such systems is therefore a graph, or network: a set 
of vertices representing the elements (genes, species, neurons, banks) connected by edges which 
capture their interactions \cite{Newman_rev,boccaletti2006complex}. Much attention has been devoted to complex networks 
over the past two decades, and one begins to discern 
an opinion forming to the effect that the fundamental properties of these constructs are now well 
understood.
This is not yet the case, however, when it comes to directed networks, or digraphs.

It is known that in many, if not perhaps most, complex systems the interactions between elements are not 
necessarily symmetric, so they are best described by directed networks (in which edges can be represented 
with arrows rather than lines).
Yet while some authors have studied this characteristic and certain of its effects
explicitly \cite{bang2008digraphs,beguerisse2014interest,qu2014nonconsensus,harush2017dynamic},
it is far more common to treat directionality as an afterthought,
as though direction were just a random binary number associated with each edge.

In fact, 
the directions of edges in a network can exhibit a degree of global order somewhat analogous to 
ferromagnetism in spin systems.
In some networks edge directions are indeed statistically independent of each other. But in others 
they can be aligned to a greater of lesser degree with a global direction. 
And there is evidence to suggest that this kind of organisation is key to understanding many 
topological and dynamical features of complex systems.

At least two strands of work on directed networks have recently uncovered some of these effects. On the one hand, 
the observation that the adjacency matrices describing empirical directed networks can be 
highly {\it non-normal} (i.e. they do not commute with their transpose) \cite{asllani2018structure}. 
On the other, that directed networks exhibit {\it trophic coherence} (that is, there exists a 
more or less well-defined
hierarchy of 
vertices, such as among plants, herbivores and carnivores in an 
ecosystem) \cite{Johnson_trophic,Johnson_looplessness}.
As I go on to show, these two features are closely related, and affect many other topological properties, 
such as whether there will exist a {\it giant strongly connected component} of vertices that are mutually reachable.
Trophic coherence has also been related to the prevalence of motifs such as feed-forward loops \cite{Janis_motifs},
and to {\it intervality}, a property associated with food webs but observed 
in other directed networks too \cite{dominguez2016intervality}.

Directionality can also have a crucial effect on dynamical systems. In previous work we have shown, for instance,
that trophic coherence is sometimes a determining factor in ecosystem stability \cite{Johnson_trophic},
or whether spreading processes such as epidemics will become endemic \cite{klaise2016neurons}.
And both trophic coherence and non-normality are reflected in graph eigenspectra,
which in turn can be related with the stability of dynamical systems \cite{Johnson_looplessness,asllani2018structure}.

I show here another example, compelling for its simplicity and richness of behaviour.
%
%
%
I simulate 
a system of binary variables, updated at every time step according to the majority rule,
on the 
neural architecture of the only fully mapped animal brain, that of the worm 
{\it C. elegans} \cite{CElegans_neural}.
Because of the worm's trophic coherence, the activity on its network is markedly different from that 
on a random graph, hopping between states where its random counterpart is stable.


The main conclusion is that there is a common origin to many of the distinctive features of certain 
directed networks: a global ordering of edge directions which leads to very different 
topological and dynamical properties than we might have expected from naïvely extending results for
undirected networks to the directed case.
However, we still have much to learn about digraphs and the effects of directionality on complex systems.


\section*{Results}

\subsection*{Trophic levels and coherence}
\noindent
Consider a directed graph with adjacency matrix $A$ (where an element $a_{ij}=1$ means there is a directed edge from 
vertex 
$v_j$ to vertex $v_i$, whereas $a_{ij}=0$ if not).
There are $N$ vertices, $L$ edges, $B$ basal vertices (i.e. vertices with no in-coming edges), and $L_B$ basal edges (edges connected
to basal vertices).
Each vertex $v_i$ has an in-degree $k_i^{in}=\sum_j a_{ij}$, and an out-degree $k_i^{out}=\sum_j a_{ji}$.

The standard definition of the trophic level of vertex $v_i$ is
\begin{equation}
s_i=1+\frac{1}{k_i^{in}}\sum_j a_{ij} s_j,
\label{eq_s}
\end{equation}
unless $v_i$ is basal (i.e. $k_i^{in}=0$), in which case $s_i=1$ by (ecological) convention \cite{Levine_levels}.
Trophic coherence is the extent to which a network is well organised into trophic levels,
and can be measured
in the following way  \cite{Johnson_trophic}.
We assign to each edge a {\it trophic difference}, $x_{ij}=s_i-s_j$. For a given network, the distribution of differences
over all edges, $p(x)$, will have mean $[ x] = 1$ and variance
$\sigma^2=[ x^2]-1$, where $[\cdot]=L^{-1}\sum_{ij}a_{ij}(\cdot)$ indicates an average over edges.
We define the {\it incoherence parameter}, $q$, as the standard deviation of $p(x)$, $q=\sigma$.
A perfectly coherent network, in which vertices fall into clearly defined trophic levels (with integer values for these)
will have $q=0$. Larger values of $q$ indicate a departure from this well-ordered state.




\begin{figure}[h!]
\begin{center}
\includegraphics[scale=0.5]{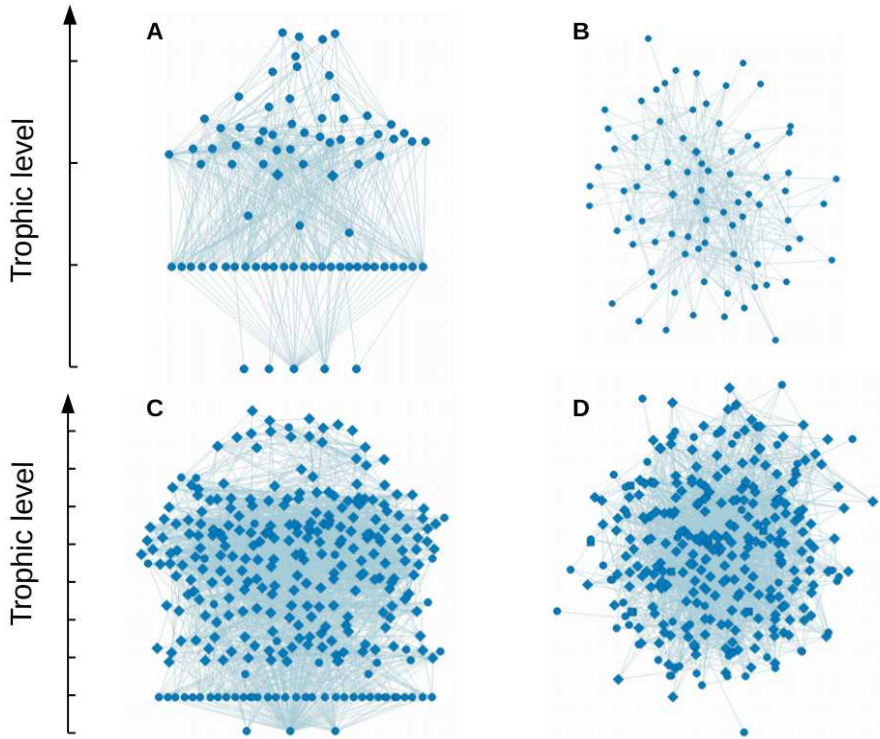}
\end{center}
\caption{
Two directed networks: the Ythan Estuary food web (A and B) and the the {\it C. elegans} neural network (C and D).
In panels A and C, the height of each vertex corresponds to its trophic level.
In panels B and D, networks are plotted according to a standard energy minimisation algorithm.
In all cases, vertices belonging to the largest strongly connected component are represented as diamonds, while the rest appear as
circles.
}
\label{fig_4nets}
\end{figure}

Eq. (\ref{eq_s}) can be written in matrix form as
\begin{equation}
\Lambda \mathbf{s}= \mathbf{z},
\label{eq_sL}
\end{equation}
where $z_i=\mbox{max}(k_i^{in},1)$ and $\Lambda=\mbox{diag}(\mathbf{z})-A$.
Each vertex can be assigned a unique trophic level if and only if $\Lambda$ is invertible.
Because the sum of elements of $\Lambda$ over any row corresponding to a non-basal vertex is zero,
$\Lambda$ will be singular for graphs with no basal vertices (i.e. there will be a right eigenvector
$\mathbf{u}=(1,1,..1)$ with eigenvalue $\lambda=0$).
Therefore, the definition of trophic levels depends on there being at least one basal vertex.

Figure \ref{fig_4nets} shows two directed networks -- the Ythan Estuary food web \cite{Ythan96} (panels A and B)
and the {\it C. elegans} neural network \cite{CElegans_neural} (C and D) -- each plotted in two different ways. 
On the left (panels A and C)
the height of each vertex corresponds to its trophic level, as indicated by the vertical axes.
For comparison, panels on the right (B and D) are plotted according to
a standard energy-minimisation method for graph visualisation: such layouts are good at highlighting 
community structure, but do not 
give any indication about the trophic structure.
These examples illustrate how the trophic level of a vertex corresponds to its position in a hierarchy.
For instance, in a food web biomass usually originates in plants
(basal or source vertices), 
flows through herbivores, then different kinds of omnivores or carnivores, 
and ends in apex predators (sinks).
Similarly, in a neural network, information enters through the sensory neurons, 
is processed through various kinds of inter-neuron, and
finally reaches the motor neurons.
Using trophic levels to determine vertex function has long been standard in ecology, but it appears likely that 
this classification would
be informative in a wide variety of complex systems describable as directed networks.

Note that the definition of trophic levels, and hence coherence, can be easily extended to the case of weighted
networks, by considering a non-binary adjacency matrix.
It is also possible to define trophic levels, and coherence, on $A^T$ instead of on $A$,
with sink vertices taking on the role of basals.
For simplicity I focus here only on unweighted networks and levels as defined by Eq. (\ref{eq_s}). 


\subsection*{Graph ensembles}
\noindent
A fruitful approach in studying random graphs is to consider ensembles,
or sets of possible networks which meet certain constraints. For instance, the 
Erd\H{o}s-R\'enyi ensemble is the set of all possible undirected networks with $N$ vertices 
and $L$ edges \cite{ErdosRenyi},
while the directed configuration ensemble comprises all directed networks with given in- and out-degree
sequences, ${\bf k^{in}}$ and ${\bf k^{out}}$ \cite{Charo}.

In Ref. \cite{Johnson_looplessness} we present results based on the {\it coherence ensemble},
which is defined as the directed configuration ensemble with the added constraint of a given trophic coherence.
We also make use of the {\it basal ensemble}, which is again based on the directed configuration ensemble, 
with the extra requisite that all non-basal vertices receive the same proportion $L_B/L$ of in-coming edges 
from 
basal vertices.
The basal ensemble is equivalent to the directed configuration ensemble 
in the limit $N\rightarrow \infty$, with $L/N\rightarrow \infty$.
Expected values of properties $y$ in these ensembles are denoted $E(y)=\overline{y}$ in the coherence ensemble, and
$E(y)=\tilde{y}$ in the basal ensemble. 
Graph ensembles such as these not only provide a powerful mathematical tool to investigate the topological 
properties of large networks; they can also be used as null models for ascertaining the extent to which 
measurements on empirical networks are statistically significant. For example, comparing the $q$ value of a network
with its basal expectation $\tilde{q}$ reveals whether it is more or less coherent than would be expected from 
its degree sequence if all else were random. Thus, in our data set, the food webs have a mean ratio 
of $q/\tilde{q}=0.44 \pm 0.17$, while for the metabolic networks it is $q/\tilde{q}=1.81 \pm 0.11$,
which implies significant coherence and incoherence, respectively, for each class (details of each network can be found in 
the Supplementary Material (SM)) \cite{Johnson_looplessness}.

In the coherence ensemble, we have shown that, in expectation,
\begin{equation}
\mbox{tr}(A^k)=\frac{\tilde{\alpha}\tilde{q}}{\alpha q}e^{\tau k},
\label{eq_tr}
 \end{equation}
where $\tau$ is the {\it loop exponent},
\begin{equation}
\tau=\ln\alpha+\frac{1}{2\tilde{q}^2}-\frac{1}{2q^2}, 
\label{eq_tau}
\end{equation}
and
$\alpha=\langle k^{in}k^{out}\rangle/\langle k\rangle$ is 
the branching factor
(the notation $\langle \cdot \rangle=N^{-1}\sum_i (\cdot)$ stands for an average over vertices).
The basal-ensemble expectations for 
$q$ and $\alpha$
  are
$\tilde{q}=\sqrt{L/L_B-1}$ and $\tilde{\alpha}=(L-L_B)/(N-B)$ \cite{Johnson_looplessness}.
%
From this, one can derive expectations for various topological properties as a function of trophic coherence.
Moreover, we have found that several kinds of empirical network -- including food webs and
examples of genetic, metabolic, neural, international trade, P2P and word adjacency networks --
conform closely to these coherence-ensemble expectations.

According to Gelfand's formula \cite{Gelfand}, the spectral radius of $A$ is
\begin{equation}
 \rho=\lim_{k\rightarrow \infty}\|A^k\|^{1/k},
\label{eq_Gelfand}
 \end{equation}
for any matrix norm $\|\cdot \|$.
Taking the norm in Eq. (\ref{eq_Gelfand}) to be the trace, we can use Eq. (\ref{eq_tr}) to find the 
expected value of the spectral radius in the coherence ensemble:
\begin{equation}
 \overline{\rho}=e^\tau.
\label{eq_rho}
 \end{equation}

The loop exponent given by Eq. (\ref{eq_tau}) can take positive or negative values. 
If $\alpha>1$ and the network in question has the trophic coherence 
of the basal ensemble
($q=\tilde{q}$), the loop exponent $\tau$ is positive
and the spectral radius is $\rho=\alpha$, as in the directed configuration ensemble.
However, if the network is sufficiently coherent ($q\rightarrow 0$), $\tau$ will be negative.
In this case, the spectral radius tends to zero: $\rho\rightarrow 0$. 
We classify networks into the {\it loopful} ($\tau>0$) and 
{\it loopless} ($\tau<0$) regimes, since the two have markedly different topological properties.
For instance, the number of directed cycles of length $\nu$ grows exponentially with $\nu$ 
in the loopful regime, while it decays exponentially to zero in the loopless one. 
I am referring here to directed cycles in which the same vertex can appear more than once (circuits), 
not to `simple cycles' in which this is not allowed.
Dom\'inguez-Garc\'ia {\it et al.} found that the simple cycles in several kinds of directed network
seem to be suppressed because of an `inherent directionality' \cite{Vir_loops}.
In Ref. \cite{Johnson_looplessness}
we discuss how this effect, too, can be explained by considering the coherence ensemble.

\subsection*{Strong connectivity} 
\noindent
A directed graph is said to be strongly connected if it is possible to reach any vertex from any other
along a directed path. It is weakly connected if this is possible when edge directions are ignored.
A weakly connected graph may have strongly connected subgraphs, and the largest of these is the 
`strongly connected component' (SSC).
Directed cycles are strongly connected subgraphs, and
large strongly connected subgraphs necessarily contain long cycles.
So 
it follows from the analysis above that in the loopless regime
the
SCC
will be vanishingly small -- 
whereas it will comprise a finite proportion of any network in the 
loopful regime.

\begin{figure}[t!]
\begin{center}
\includegraphics[scale=0.5]{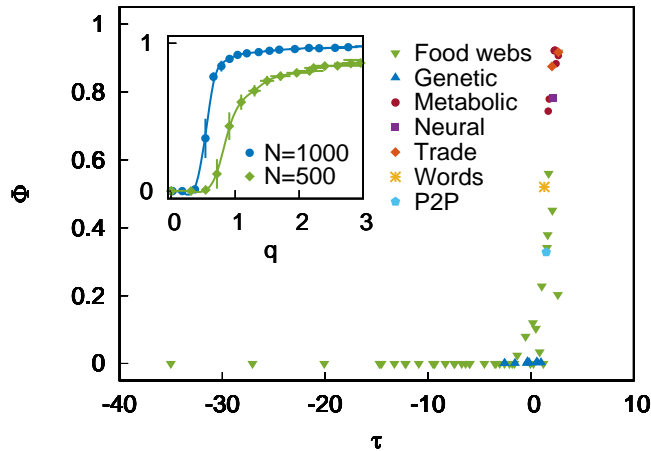}
\end{center}
\caption{
Fraction of non-basal vertices in the strongly connected component, $\Phi$, against loop exponent, $\tau$, 
for several empirical networks. Inset: $\Phi$ against $q$ for networks generated with 
the `preferential preying model' \cite{klaise2016neurons}.
}
\label{fig_web}
\end{figure}

In Fig. \ref{fig_4nets}, vertices belonging to the 
SCC, in each network, 
are represented with diamonds, in contrast to the circles used for other vertices.
In the food web ($\tau=-1.32$) the SCC only has two vertices, while in the neural network ($\tau=2.17$)
the SCC includes most of the vertices.
Figure \ref{fig_web} shows the size of the SCC as a fraction of the number of non-basal vertices, 
$\Phi$, against 
$\tau$ for several empirical networks of various kinds.
We observe that $\Phi\simeq 0$ when $\tau<0$, and $\Phi> 0$ when $\tau>0$.
Details of each network can be found in the SM.
This disparity in strong connectivity could not be understood by simply extending known results for 
undirected networks, according to which the main factor determining $\Phi$ 
is the mean degree, $\langle k\rangle$ \cite{Newman_rev}.
Here we see some networks, such as most of the metabolic ones (Table S3 of SM), with 
$\Phi>0.9$ and $\langle k\rangle < 3$; whereas many of the food webs (Table S1 of SM) have
 $\Phi\simeq 0$ despite being much denser ($\langle k \rangle >10$).
It is only by measuring their trophic coherence, and hence $\tau$, that the reason for this 
becomes clear.

Networks with specified trophic coherence can be generated computationally with
the `preferential preying model' \cite{klaise2016neurons}, which is described below in 
Methods \ref{sec_PPM}.
The inset in Fig. \ref{fig_web} shows how $\Phi$ varies with $q$ in networks simulated with this model.
The model displays what appears to be a continuous (i.e. second order) phase transition in strong connectivity 
with coherence, reminiscent of 
the percolation transition observed in undirected, Erd\H{o}s-R\'enyi random graphs 
with mean degree
\cite{Newman_rev}.

\subsection*{Non-normality}
\noindent
An $N\times N$ matrix $A$ is said to be {\it normal} if its adjacency matrix, $A$, commutes with its 
transpose, $AA^T = A^TA$;
or, conversely, it is {\it non-normal} if $AA^T\neq A^TA$ \cite{asllani2018structure}.
Clearly, if $A$ is the adjacency matrix of a network, it must be directed to be non-normal.
Intuitively, we might expect a large
deviation from normality to indicate a network with a well-defined directionality.
This is also what occurs in trophically coherent networks.

Asllani {\it et al.} have recently shown that a wide variety of empirical networks are highly non-normal,
and they discuss the significant implications for dynamical systems with such a structure
\cite{asllani2018structure}.
To quantify this property they use
Hermici's departure from normality: 
\begin{equation}
 D_F=\sqrt{\|A\|_F^2-\sum_i|\lambda_i|^2},
\end{equation}
where
\begin{equation}
\|A\|_F=\sqrt{\sum_{ij}|a_{ij}|^2}
\end{equation}
is the Frobenius norm \cite{Trefethen}.
And, in order to compare matrices of different sizes, 
they use the normalised version:
\begin{equation}
 d_F=\frac{D_F}{\|A\|_F}.
\end{equation}
A normal matrix will have $d_F=0$, and $d_F$ is closer to $1$ the more $A$ departs from normality.

\begin{figure}[ht!]
\begin{center}
\includegraphics[scale=0.5]{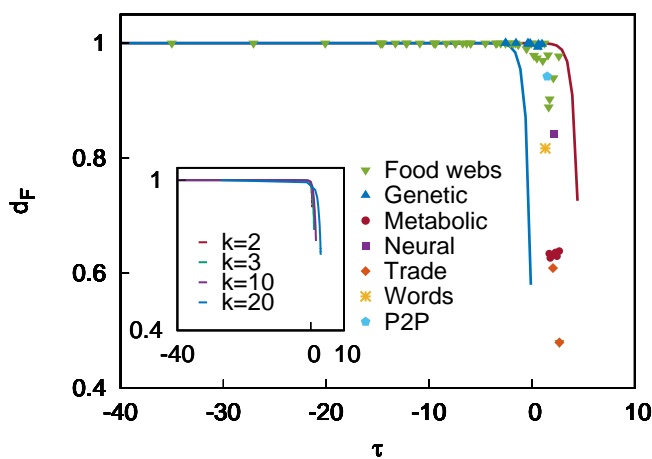}
\end{center}
\caption{
Normalised deviation from normality, $d_F$, against loop exponent, $\tau$, for several empirical networks.
The lower (blue) line is Eq. (\ref{eq_dFL}) for $\langle k\rangle =1.23$;
the upper (red) line is Eq. (\ref{eq_dFU}) for $L=15317$ (these are the lowest $\langle k\rangle$ and highest $L$,
respectively, found in the set of networks).
The inset shows Eq. (\ref{eq_dFL}) for various different mean degrees: $\langle k\rangle=2$, $3$, $10$, and $20$.
}
\label{fig_dev}
\end{figure}

\subsection*{Trophic coherence and non-normality}
\noindent
Using results for the coherence ensemble, it is possible to relate trophic coherence with non-normality. In particular, 
the following theorem holds in this ensemble:
\\

\noindent
{\bf Theorem.}
{\it 
The expected deviation from normality, $\overline{d_F}$, for digraphs drawn from 
the coherence ensemble tends to $1$ with increasing trophic coherence. That is,}
\begin{equation}
 \lim_{q\rightarrow 0}\overline{d_F}=1.
\end{equation}
{\it Furthermore,} 
\begin{equation}
  \overline{d_F}>\sqrt{1-\frac{1}{\langle k\rangle}}
\end{equation}
{\it for digraphs in the $\tau<0$ regime, where
$\langle k\rangle$
is the mean degree.}
\\


\noindent
{\bf Proof.}
For a binary (i.e. unweighted) adjacency matrix $A$ we have $\|A\|_F^2=L$, where $L$ is the number of edges.
So we can express the normalised departure from normality as
\begin{equation}
 d_F=\sqrt{1-\frac{1}{L}\sum_i|\lambda_i|^2}.
\end{equation}





Let $\rho=\rho(A)$ be the spectral radius of $A$.
Since $|\lambda_i|\leq \rho$ for all $i$, 
and $|\lambda_j| = \rho$ for at least one $j$,
we have that 
\begin{equation}
\rho^2\leq \sum_i^N |\lambda_i|^2\leq N\rho^2.
\end{equation}
We can use these bounds on $\sum_i^N |\lambda_i|^2$ to define a lower bound, $d_F^L$, and an upper bound,  $d_F^U$, 
on $d_F$ in terms of the spectral radius $\rho$:
\begin{equation}
 d_F^L=\sqrt{1-\frac{N}{L}\rho^2},
\label{eq_dl}
\end{equation}
\begin{equation}
 d_F^U=\sqrt{1-\frac{1}{L}\rho^2}.
\label{eq_du}
\end{equation}
Inserting Eq. (\ref{eq_rho}) into Eqs. (\ref{eq_dl}) and (\ref{eq_du}) provides lower and upper bounds on the expected deviation from normality
in the coherence ensemble:
\begin{equation}
 \overline{d_F^L} = \sqrt{1-\frac{N}{L}e^{2\tau}},
\label{eq_dFL}
 \end{equation}
\begin{equation}
 \overline{d_F^U} = \sqrt{1-\frac{1}{L}e^{2\tau}}.
\label{eq_dFU}
 \end{equation}

Eq. (\ref{eq_dFL}) implies that 
\begin{equation}
 \lim_{\tau\rightarrow -\infty} \overline{d_F} = \lim_{\tau\rightarrow -\infty} \overline{d_F^L} = 1.
\end{equation}
Moreover, in the loopless regime ($\tau<0$) we have 
\begin{equation}
\overline{d_F}\geq \overline{d_F^L}(\tau=0)=\sqrt{1-\frac{1}{\langle k\rangle}}.
\end{equation}
$\square$ 

In other words, {\it coherent networks are highly non normal.}

%
%
%
%

Figure \ref{fig_dev} shows the normalised deviation from normality, $d_F$, against the loop exponent, $\tau$, 
for the same set of empirical 
networks used in Fig. \ref{fig_web}. The blue line is $\overline{d_F^L}$ as given by 
Eq. (\ref{eq_dFL}) for the case of the network with lowest mean 
degree in the set; while the red line is $\overline{d_F^U}$ as given by 
Eq. (\ref{eq_dFU}) for the network with the largest number of edges.
The inset shows Eq. (\ref{eq_dFL}) for various different mean degrees.
We can observe that the real networks with small or negative $\tau$ are indeed highly non-normal, 
and the bounds obtained for the coherence ensemble
hold for these empirical cases too.

\subsection*{Dynamical stability}
\noindent
We have shown in previous work that trophic coherence can have an important bearing on the dynamics of complex systems.
In particular, it affects linear stability in food-web models \cite{Johnson_trophic}, 
and percolation in spreading processes \cite{klaise2016neurons}.
I go on to show another example in which trophic coherence has a remarkable effect on the stability of one of the simplest
dynamical systems.

Consider a set of variables on the vertices of a network which can, at each discrete time step $t$, take either of two states,
$\sigma_i(t)=\pm 1$, according to the `majority rule' applied to the states of in-neighbours. In other words,
if $h_i(t)=\sum_j a_{ij}\sigma_j(t)$, then $\sigma_i(t+1)= +1$ if $h_i(t)>0$, and $\sigma_i(t+1)= -1$ if $h_i(t)<0$.
If $h_i(t)=0$, one of the two states is chosen randomly with equal probability
(this is the only source of stochasticity in the dynamics).
The variables are updated in parallel at each $t$, and
the overall state of the system can be measured with the mean activity, 
$m(t)=\langle \sigma(t)\rangle$.
This dynamics is a version of the `majority rule' model used as a simple approximation to opinion formation \cite{krapivsky2003dynamics}, 
and coincides with a Hopfield neural network model when all synaptic weights are equal,
and with a zero-temperature Ising model on a directed network \cite{marro2005nonequilibrium}.

\begin{figure}[th!]
\begin{center} 
\includegraphics[scale=0.35]{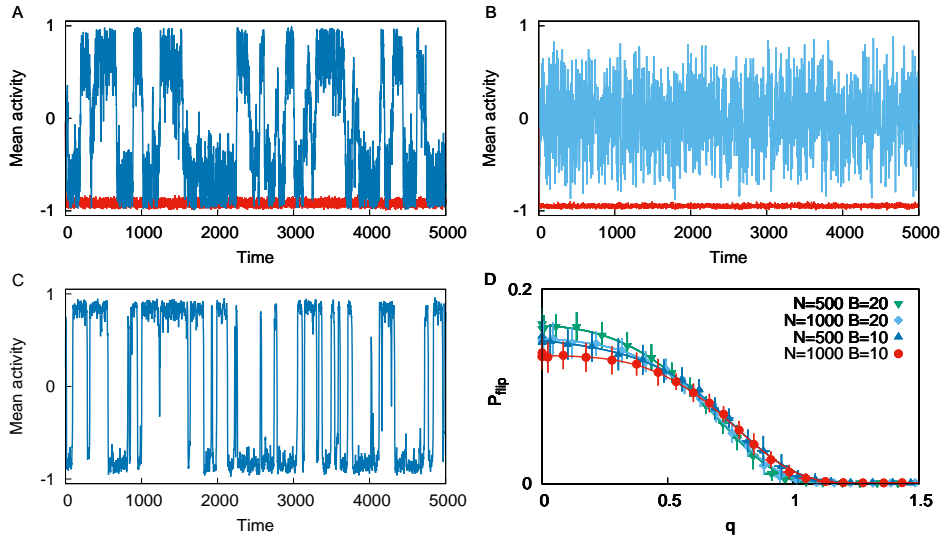}
\end{center}
\caption{
{\bf A}: Time series from simulations of the majority rule dynamics (described in the main text)
on the {\it C. elegans} neural network (blue) and a randomisation of this network which preserves degree sequences (red).
{\bf B}: Majority rule dynamics on two networks generated with the preferential preying model and the same $N$, $B$ and $L$ as in 
panel A, but with different trophic coherence, set by the parameter $T$: $T=0.01$, which leads to $q\simeq 0$ (cyan),
and $T=10$, yielding $q\simeq \tilde{q}$ (red).
{\bf C}: As in panel B, but with $T=1$, which leads to a similar $q$ to the neural network.
{\bf D}: Frequency with which the sign of $m(t)$ changes in majority rule dynamics against $q$, when run on preferential preying model
networks for different values of $N$ and $B$ ($\langle k\rangle=10$ in all cases).
}
\label{fig_cel_P}
\end{figure}
%

Figure \ref{fig_cel_P}.A shows a time series of $m(t)$ obtained from simulations of this model on two different networks. 
The blue line is for the neural network of the worm {\it C. elegans} -- i.e. the same network which is represented in 
panels C and D of Fig. \ref{fig_4nets}. The red line is for a randomisation of this same network, achieved by repeatedly 
choosing two pairs of connected vertices, and swapping the out-neighbours. This randomisation preserves the in- and out-degrees 
of every node while destroying other structure.
Activity on the randomised network is stable, with the mean activity remaining in this case close to $-1$ (other simulations are 
equally likely to adopt $m(t)\simeq +1$).
However, on the empirical network the mean activity switches between positive and negative states.
This instability must be caused by some topological difference between the two networks other than the degree sequences.

Figure \ref{fig_cel_P}.B shows the same dynamics but on two networks generated 
with the `preferential preying model' \cite{klaise2016neurons}, using the same numbers of nodes, basal nodes and edges 
as the empirical network has. The red line is for a network with the coherence of a random graph
($q\simeq \tilde{q}$), and the cyan line for a highly coherent one ($q\simeq 0$).
The incoherent network presents stable dynamics, like the randomised version of the neural network; while the 
coherent one is completely unstable, with $m(t)$ changing sign every few time steps.
Figure \ref{fig_cel_P}.C shows $m(t)$ on a network
with
intermediate trophic coherence, similar to the 
neural network's value ($q/\tilde{q}\simeq 0.4$).  
Now we recover the bistability of the empirical topology, which suggests that it is indeed trophic coherence which 
accounts for this dynamical behaviour.

Figure \ref{fig_cel_P}.D shows average results of such simulations on preferential preying networks with different numbers of 
nodes and basal nodes. The proportion of time steps in which $m(t)$ changes sign, $P_{flip}$, is plotted against $q$, revealing 
what appears to be a continuous transition between and unstable and a stable phase, at $q\simeq 1$.
Intuitively, we can understand this behaviour by considering that the basal vertices always have $h_i=0$, and so take either 
state $+1$ or $-1$ with a $1/2$ probability at each $t$. In a maximally coherent network, this random configuration
propagates up through the levels, leading to fully unstable behaviour.
On the other hand, when the network is highly incoherent, with a large strongly connected component,
most vertices are less susceptible to the influence of the basal vertices, and tend instead to preserve the average 
state of the network.

It is possible to carry out a mean-field approximation for networks drawn from the basal ensemble, which reveals a critical 
$q$ separating stability from instability at $q_c=1$ (this is done in Methods \ref{sec_maj_rule}).
Although this is not such a straightforward exercise for the coherence ensemble, it appears from numerical simulations that 
the same critical value may also apply to such networks (see Fig. \ref{fig_cel_P}.D).

The fact that the emergence of bistability stems from the stochasticity of the basal vertices might seem 
like an artefact of this toy model. But there are many networks
in which these vertices may represent 
sources of information -- e.g. sensory neurons, oligarchs, oil-producing nations -- 
which cascades through the system. These results suggest that how the system reacts to novel 
information entering through the basal vertices depends strongly on trophic coherence.

\section*{Concluding Remarks}

%

\noindent
Directed networks can exhibit topological properties which are much more than 
trivial extensions of the features available to their undirected counterparts.
Edges can be organised according to a global direction in a way analogous to 
the alignment of spins in a ferromagnet.
Some hallmarks of this phenomenon have recently been identified, such as 
trophic coherence \cite{Johnson_trophic} and non-normality \cite{asllani2018structure}.
One of the results in this paper is that these two properties are closely related.

Directed networks can belong to 
either of two regimes \cite{Johnson_looplessness}. In the `loopful' regime edges are not strongly aligned with a global direction. 
This manifests in topologies that are trophically incoherent, with small deviations from normality,
large spectral radii, large strongly connected components, and numbers of 
cycles which grow exponentially with length.
In the `loopless' regime, however,
edges are organised according to a global direction.
This translates into all the above properties being inverted:
networks are highly coherent and non-normal,
spectral radii and strongly connected components are vanishing, 
and numbers of cycles decay exponentially with length.
All these properties can be related, at least in expectation, to the `loop exponent' $\tau$, 
which is a function only of trophic coherence and the in- and out-degree sequences.
The sign of $\tau$ determines which regime a network belongs to.

These topological features can have an important influence on the behaviour of dynamical systems 
in which the interactions between elements are not symmetric. In previous work we have shown that 
trophic coherence increases linear stability in ecological models, and provides a possible solution 
to May's paradox -- i.e. the fact that large ecosystems seem to be more stable than smaller 
ones \cite{May,Johnson_trophic}. Asllani {\it et al.} have also related non-normality with 
stability \cite{asllani2018structure}.
In the case of spreading processes like models of epidemics, trophic coherence can determine whether
activity (e.g. an infection) dies out quickly or becomes endemic \cite{klaise2016neurons}.

The example in this paper shows that the relationship between coherence and stability is not straightforward,
and that quite rich and unexpected behaviour can emerge even in a very simple dynamical model. 
A system of binary elements updated according to the majority rule is stable on incoherent networks, as 
would also be the case on an undirected network. However, on a highly coherent network the system is 
completely unstable. And in the case of an intermediate coherence the mean activity hops between metastable states, 
with switching times that depend on coherence.
Given that many real-world networks -- such as the {\it C. elegans} neural network used here for 
illustration -- have intermediate levels of coherence, this may be an important effect in a wide variety of 
complex, dynamical systems.

These results suggest that a complete understanding of the relationship between structure and 
function in complex systems will require further fundamental research, in particular as regards topological 
properties related to edge directionality.
While it is possible to tie together several of these properties -- 
such as trophic coherence, non-normality and strong connectivity -- we have not yet considered 
how these might interact with other topological characteristics, like degree distributions, community structure or 
assortativity \cite{Newman_rev,johnson2010entropic}.
And given the highly disparate kinds of dynamical behaviour we have seen even between quite simple models,
it is clear that a more exhaustive investigation is needed.

Two tools which could be of use in such an endeavour are the {\it preferential preying model}, which
provides a way of generating networks with specified coherence numerically \cite{Johnson_trophic,klaise2016neurons};
and the {\it coherence ensemble}, a theoretical approach which allows one to investigate the relationships between 
properties mathematically \cite{Johnson_looplessness}.



\section*{Methods}

\subsection{The preferential preying model}
\label{sec_PPM}

%

We can generate networks with a given trophic coherence with the model used in Ref. \cite{klaise2016neurons},
which is a generalisation of the one first proposed in Ref. \cite{Johnson_trophic}. (The original version is loosely 
inspired by immigration of species into an ecosystem, and somewhat resembles Barab{\'a}si and Albert's famous preferential
attachment model \cite{Newman_rev} -- hence the name.)

We begin with $B$ basal
vertices and proceed to introduce 
$N-B$ non-basal vertices sequentially.
Each vertex is initially assigned a single in-neighbour, chosen randomly
from among the vertices (basal and non-basal) already in the network when it arrives.
At this stage vertex $v_i$ has a preliminary trophic level $s_i'$, as given by Eq. (\ref{eq_s})
(note that this simply means assigning $s_i'=s_j'+1$ if $v_i$ is given $v_j$ as 
its in-neighbour).

We then introduce
the remaining $L-N+B$ edges needed to make up a total of $L$.
For this, each pair of nodes $\lbrace v_i,v_j \rbrace$ 
such that $v_i$ is a non-basal vertex
is attributed a temporary trophic distance $\tilde{x}_{ij}'=\tilde{s}_i'-\tilde{s}_j'$. 
Edges between pairs are then placed with
a probability proportional to
\begin{equation}
P(a_{ij}= 1)\propto \exp\left[-\frac{(x_{ij}'-1)^2}{2T^2}\right],
\label{eq_Plink}
\end{equation}
until there are $L$ edges in the network.
The `temperature' parameter $T$ sets the network's
trophic coherence,
with $T=0$ yielding maximally coherent networks ($q=0$), and incoherence increasing monotonically with $T$.
The specific choice for the edge probability is arbitrary, but the form in Eq. (\ref{eq_Plink}) is conducive to
a Gaussian distribution of distances $x$, which we have found to be a good fit to empirical data on several kinds of
networks.

Unless a maximally coherent network is intended, one must then recalculate the actual trophic levels ${\bf s}$
of the final networks, according to Eq. (\ref{eq_sL}), and measure $q$ based on these.
Although in practice $q$ will not generally be equal to $T$, one can easily obtain, 
for given $\lbrace N, B, L \rbrace$, the value of 
$T$ which best approximates the intended $q$, thanks to the monotonic relation 
between $T$ and $q$ \cite{Johnson_trophic,klaise2016neurons}. 


\subsection{Majority rule dynamics on basal ensemble networks}
\label{sec_maj_rule}

\noindent
Consider the majority rule dynamics described above on networks drawn from the basal ensemble \cite{Johnson_looplessness}.
In such networks all non-basal vertices have the same proportion of in-coming edges from basal nodes.
In a mean-field approximation, 
the field at any non-basal vertex $i$ is
\begin{equation}
 h_i=\sum_j a_{ij}s_j =  k_i^n m^n + k_i^b m^b,
\end{equation}
where $k_i^b$ and $k_i^n$ are the mean numbers of incoming edges from basal and non-basal neighbours, respectively;
and $m^b$ and $m^n$ are the mean activities of basal and non-basal vertices, respectively (time dependencies have been dropped for clarity).
If a non-basal vertex $v_i$ is in the state $s_i=\mbox{sgn}(m^n)$ at time $t$, the probability that it will change state at $t+1$
will be
\begin{equation}
 P_{flip}= \mbox{Pr}(m^n m^b <0) \mbox{Pr}( k_i^n |m^n| < k_i^b |m^b|)
\end{equation}
(where $\mbox{Pr}(y)$ stands for the probability of event $y$).
Because the network is drawn from the basal ensemble, we have that
\begin{equation}
 \frac{k_i^n}{k_i^b}=\frac{L}{L_B}-1\equiv \lambda.
\end{equation}
At each time step,
every basal vertex's state will be $+1$ or $-1$ with equal probability. Therefore,
the number of basal vertices in the $+1$ state, $n_{+}$, 
will be a random draw from a binomial distribution:
\begin{equation}
n_{+} \sim \mbox{Bin}(n=B; p=1/2),
\end{equation}
where 
\begin{equation}
n_{+}=\frac{1}{2}B(m^b+1). 
\end{equation}


Let us consider the case in which the majority of non-basal vertices are in the same state, so that $|m^n|\simeq 1$;
and, without loss of generality, that $m^n<0$.
We now have
\begin{equation}
 P_{flip}= \mbox{Pr}( \lambda < m^b) = \mbox{Pr}[n_{+}> (\lambda+1)B/2].
 \end{equation}
The CDF of the binomial distribution $\mbox{Bin}(n,p)$ is given by the incomplete 
Beta function:
$\mbox{Pr}(X\leq k)=\mbox{I}_{1-p}(n-k,1+k)$.
Therefore, we have
\begin{equation}
 P_{flip}= 1-\mbox{I}_{1/2}\left[\frac{1}{2}(1-\lambda)B,1+\frac{1}{2}(1+\lambda)B\right].
 \label{eq_Pflip}
 \end{equation}
One corollary of Eq. (\ref{eq_Pflip}) is that $P_{flip}>0$ requires $\lambda<1$, or $L<2L_B$.

The symmetry of the system implies that $P_{flip}$ is independent of the sign of $m^n$. Therefore, 
$m^n$ will follow a Bernoulli process with $p=P_{flip}$.
The distribution of time intervals,
$\Delta$, 
between sign changes of $m^n$ will therefore follow
\begin{equation}
 P(\Delta)=P_{flip}(1-P_{flip})^{\Delta-1}.
 \label{eq_Ptau}
\end{equation}




%
In the basal ensemble, we have
$$
\tilde{q}=\sqrt{\frac{L}{L_B}-1}
$$
Therefore, the critical ratio $L/L_B=2$ obtained above
is equivalent to a critical coherence
$$
\tilde{q}_c=1.
$$

This mean-field analysis is only valid for the basal ensemble, but simulations of the preferential preying model suggest that 
$q_c\simeq 1$ applies to other kinds of networks too (see Fig. \ref{fig_cel_P}D).


%

\begin{thebibliography}{10}

\bibitem{Newman_rev}
M.~E.~J. Newman, ``The structure and function of complex networks,'' {\em SIAM
  Review}, vol.~45, pp.~167--256, 2003.

\bibitem{boccaletti2006complex}
S.~Boccaletti, V.~Latora, Y.~Moreno, M.~Chavez, and D.-U. Hwang, ``Complex
  networks: Structure and dynamics,'' {\em Physics Reports}, vol.~424, no.~4-5,
  pp.~175--308, 2006.

\bibitem{bang2008digraphs}
J.~Bang-Jensen and G.~Z. Gutin, {\em Digraphs: theory, algorithms and
  applications}.
\newblock Springer Science \& Business Media, 2008.

\bibitem{beguerisse2014interest}
M.~Beguerisse-D{\'\i}az, G.~Garduno-Hern{\'a}ndez, B.~Vangelov, S.~N. Yaliraki,
  and M.~Barahona, ``Interest communities and flow roles in directed networks:
  the twitter network of the uk riots,'' {\em Journal of the Royal Society
  Interface}, vol.~11, no.~101, p.~20140940, 2014.

\bibitem{qu2014nonconsensus}
B.~Qu, Q.~Li, S.~Havlin, H.~E. Stanley, and H.~Wang, ``Nonconsensus opinion
  model on directed networks,'' {\em Physical Review E}, vol.~90, no.~5,
  p.~052811, 2014.

\bibitem{harush2017dynamic}
U.~Harush and B.~Barzel, ``Dynamic patterns of information flow in complex
  networks,'' {\em Nature communications}, vol.~8, no.~1, p.~2181, 2017.

\bibitem{asllani2018structure}
M.~Asllani, R.~Lambiotte, and T.~Carletti, ``Structure and dynamical behavior
  of non-normal networks,'' {\em Science Advances}, vol.~4, no.~12,
  p.~eaau9403, 2018.

\bibitem{Johnson_trophic}
S.~Johnson, V.~Dom\'inguez-Garc\'ia, L.~Donetti, and M.~A. Mu\~noz, ``Trophic
  coherence determines food-web stability,'' {\em Proc. Natl. Acad. Sci. USA},
  vol.~111, no.~50, pp.~17923--17928, 2014.

\bibitem{Johnson_looplessness}
S.~Johnson and N.~S. Jones, ``Looplessness in networks is linked to trophic
  coherence,'' {\em Proc. Natl. Acad. Sci. USA}, vol.~114, no.~22,
  pp.~5618--5623, 2017.

\bibitem{Janis_motifs}
J.~Klaise and S.~Johnson, ``The origin of motif families in food webs,'' {\em
  Sci. Rep.}, vol.~7, p.~16197, 2017.

\bibitem{dominguez2016intervality}
V.~Dom{\'\i}nguez-Garc{\'\i}a, S.~Johnson, and M.~A. Mu{\~n}oz, ``Intervality
  and coherence in complex networks,'' {\em Chaos}, vol.~26, no.~065308, 2016.

\bibitem{klaise2016neurons}
J.~Klaise and S.~Johnson, ``From neurons to epidemics: How trophic coherence
  affects spreading processes,'' {\em Chaos}, vol.~26, no.~065310, 2016.

\bibitem{CElegans_neural}
J.~G. White, E.~Southgate, J.~N. Thompson, and S.~Brenner, ``The structure of
  the nervous system of the nematode caenorhabditis elegans,'' {\em Phil.
  Trans. R. Soc. London}, vol.~314, pp.~1--340, 1986.

\bibitem{Levine_levels}
S.~Levine, ``Several measures of trophic structure applicable to complex food
  webs,'' {\em J. Theor. Biol.}, vol.~83, pp.~195--207, 1980.

\bibitem{Ythan96}
M.~Huxham, S.~Beaney, and D.~Raffaelli, ``{Do parasites reduce the chances of
  triangulation in a real food web?},'' {\em Oikos}, vol.~76, pp.~284--300,
  1996.
  
\bibitem{ErdosRenyi}
P.~Erd\H{o}s and A.~R\'enyi, ``On random graphs. I,'' {\em Publicationes
  Mathematicae}, vol.~6, pp.~290--297, 1959.

\bibitem{Charo}
H.~Kim, C.~I. {Del Genio}, K.~E. Bassler, and Z.~Toroczkai, ``Constructing and
  sampling directed graphs with given degree sequences,'' {\em New J. Phys.},
  vol.~14, p.~023012, 2012.
  
\bibitem{Gelfand}
I.~Gelfand, ``Normierte Ringe,'' {\em Rec. Math. [Mat. Sbornik] N.S.}, vol.~9,
  no.~51, pp.~3--24, 1941.

\bibitem{Vir_loops}
V.~Dom\'inguez-Garc\'ia, S.~Pigolotti, and M.~A. Mu{\~n}oz, ``Inherent
  directionality explains the lack of feedback loops in empirical networks,''
  {\em Sci. Rep.}, vol.~4, p.~7497, 2014.

\bibitem{Trefethen}
L.~N. Trefethen and M.~Embree, {\em Spectra and Pseudospectra: The Behavior of
  Nonnormal Matrices and Operators}.
\newblock Princeton: Princeton Univ. Press, 2005.

\bibitem{krapivsky2003dynamics}
P.~L. Krapivsky and S.~Redner, ``Dynamics of majority rule in two-state
  interacting spin systems,'' {\em Physical Review Letters}, vol.~90, no.~23,
  p.~238701, 2003.

\bibitem{marro2005nonequilibrium}
J.~Marro and R.~Dickman, {\em Nonequilibrium phase transitions in lattice
  models}.
\newblock Cambridge University Press, 2005.

\bibitem{May}
R.~M. May, ``Will a large complex system be stable?,'' {\em Nature}, vol.~238,
  pp.~413--14, 1972.

\bibitem{johnson2010entropic}
S.~Johnson, J.~J. Torres, J.~Marro, and M.~A. Munoz, ``Entropic origin of
  disassortativity in complex networks,'' {\em Physical Review Letters},
  vol.~104, no.~10, p.~108702, 2010.

\end{thebibliography}

\begin{thebibliography}{10}

\bibitem{Johnson_looplessness}
S.~Johnson and N.~S. Jones, ``Looplessness in networks is linked to trophic
  coherence,'' {\em Proc. Natl. Acad. Sci. USA}, vol.~114, no.~22,
  pp.~5618--5623, 2017.

\bibitem{benguela}
P.~Yodzis, ``Local trophodynamics and the interaction of marine mammals and
  fisheries in the benguela ecosystem,'' {\em Journal of Animal Ecology},
  vol.~67, no.~4, pp.~635--658, 1998.

\bibitem{streams5}
R.~M. Thompson and C.~R. Townsend, ``Impacts on stream food webs of native and
  exotic forest: An intercontinental comparison,'' {\em Ecology}, vol.~84,
  pp.~145--161, 2003.

\bibitem{streams6}
R.~M.~T. andC. R.~Townsend, ``Energy availability, spatial heterogeneity and
  ecosystem size predict food-web structure in stream,'' {\em Oikos}, vol.~108,
  p.~137–148, 2005.

\bibitem{streams7}
Townsend, Thompson, McIntosh, Kilroy, Edwards, and Scarsbrook, ``Disturbance,
  resource supply, and food-web architecture in streams,'' {\em Ecology
  Letters}, vol.~1, no.~3, pp.~200--209, 1998.

\bibitem{bridge}
K.~Havens, ``Scale and structure in natural food webs,'' {\em Science},
  vol.~257, no.~5073, pp.~1107--1109, 1992.

\bibitem{canton}
Townsend, Thompson, McIntosh, Kilroy, Edwards, and Scarsbrook, ``Disturbance,
  resource supply, and food-web architecture in streams,'' {\em Ecology
  Letters}, vol.~1, no.~3, pp.~200--209, 1998.

\bibitem{reef}
S.~Opitz, ``Trophic interactions in {C}aribbean coral reefs,'' {\em ICLARM
  Tech. Rep.}, vol.~43, p.~341, 1996.

\bibitem{caribbean_2005}
J.~Bascompte, C.~Melián, and E.~Sala, ``Interaction strength combinations and
  the overfishing of a marine food web,'' {\em Proceedings of the National
  Academy of Sciences of the United States of America}, vol.~102, no.~15,
  pp.~5443--5447, 2005.

\bibitem{chesapeake1}
R.~E. Ulanowicz and D.~Baird, ``Nutrient controls on ecosystem dynamics: the
  chesapeake mesohaline community,'' {\em Journal of Marine Systems}, vol.~19,
  no.~1–3, pp.~159 -- 172, 1999.

\bibitem{chesapeake2}
L.~G. Abarca-Arenas and R.~E. Ulanowicz, ``The effects of taxonomic aggregation
  on network analysis,'' {\em Ecological Modelling}, vol.~149, no.~3, pp.~285
  -- 296, 2002.

\bibitem{coachella}
G.~Polis, ``Complex trophic interactions in deserts: an empirical critique of
  food-web theory,'' {\em Am. Nat.}, vol.~138, pp.~123--125, 1991.

\bibitem{el_verde}
R.~B. Waide and W.~B.~R. ({\it eds.}), {\em The Food Web of a Tropical
  Rainforest}.
\newblock Chicago: University of Chicago Press, 1996.

\bibitem{little_rock}
N.~D. Martinez, ``Artifacts or attributes? {E}ffects of resolution on the
  {L}ittle {R}ock {L}ake food web,'' {\em Ecol. Monogr.}, vol.~61,
  pp.~367--392, 1991.

\bibitem{lough_hyne_1}
J.~Riede, U.~Brose, B.~Ebenman, U.~Jacob, R.~Thompson, C.~Townsend, and
  T.~Jonsson, ``Stepping in {E}lton's footprints: a general scaling model for
  body masses and trophic levels across ecosystems,'' {\em Ecology Letters},
  vol.~14, pp.~169--178, 2011.

\bibitem{lough_hyne_2}
A.~Eklöf, U.~Jacob, J.~Kopp, J.~Bosch, R.~Castro-Urgal, B.~Dalsgaard,
  N.~Chacoff, C.~deSassi, M.~Galetti, P.~Guimaraes, S.~Lomáscolo,
  A.~Martín~González, M.~Pizo, R.~Rader, A.~Rodrigo, J.~Tylianakis,
  D.~Vazquez, and S.~Allesina, ``The dimensionality of ecological networks,''
  {\em Ecology Letters}, vol.~16, pp.~577--583, 2013.

\bibitem{shelf}
J.~Link, ``Does food web theory work for marine ecosystems?,'' {\em Mar. Ecol.
  Prog. Ser.}, vol.~230, pp.~1--9, 2002.

\bibitem{broom}
J.~Memmott, N.~D. Martinez, and J.~E. Cohen, ``Predators, parasitoids and
  pathogens: species richness, trophic generality and body sizes in a natural
  food web,'' {\em J. Anim. Ecol.}, vol.~69, pp.~1--15, 2000.

\bibitem{skipwith}
P.~H. Warren, ``Spatial and temporal variation in the structure of a freshwater
  food web,'' {\em Oikos}, vol.~55, pp.~299--311, 1989.

\bibitem{st_marks}
R.~R. Christian and J.~J. Luczkovich, ``Organizing and understanding a winter's
  {S}eagrass foodweb network through effective trophic levels,'' {\em Ecol.
  Model.}, vol.~117, pp.~99--124, 1999.

\bibitem{st_martin}
L.~Goldwasser and J.~A. Roughgarden, ``Construction of a large {C}aribbean food
  web,'' {\em Ecology}, vol.~74, pp.~1216--1233, 1993.

\bibitem{stony}
Townsend, Thompson, McIntosh, Kilroy, Edwards, and Scarsbrook, ``Disturbance,
  resource supply, and food-web architecture in streams,'' {\em Ecology
  Letters}, vol.~1, no.~3, pp.~200--209, 1998.

\bibitem{grass}
N.~D. Martinez, B.~A. Hawkins, H.~A. Dawah, and B.~P. Feifarek, ``Effects of
  sampling effort on characterization of food-web structure,'' {\em Ecology},
  vol.~80, p.~1044–1055, 1999.

\bibitem{weddell_sea}
U.~Jacob, A.~Thierry, U.~Brose, W.~Arntz, S.~Berg, T.~Brey, I.~Fetzer,
  T.~Jonsson, K.~Mintenbeck, C.~Möllmann, O.~Petchey, J.~Riede, and J.~Dunne,
  ``The role of body size in complex food webs,'' {\em Advances in Ecological
  Research}, vol.~45, pp.~181--223, 2011.

\bibitem{Ythan96}
M.~Huxham, S.~Beaney, and D.~Raffaelli, ``{Do parasites reduce the chances of
  triangulation in a real food web?},'' {\em Oikos}, vol.~76, pp.~284--300,
  1996.

\bibitem{human_gerstein}
M.~B. Gerstein, A.~Kundaje, M.~Hariharan, S.~G. Landt, K.-K. Yan, C.~Cheng,
  X.~J. Mu, E.~Khurana, J.~Rozowsky, R.~Alexander, {\em et~al.}, ``Architecture
  of the human regulatory network derived from encode data,'' {\em Nature},
  vol.~489, no.~7414, pp.~91--100, 2012.

\bibitem{BQS}
L.~Albergante, J.~J. Blow, and T.~J. Newman, ``Buffered {Q}ualitative
  {S}tability explains the robustness and evolvability of transcriptional
  networks,'' {\em eLife}, vol.~3, p.~e02863, 2014.

\bibitem{ecoli_salgado}
H.~Salgado, M.~Peralta-Gil, S.~Gama-Castro, A.~Santos-Zavaleta,
  L.~Mu{\~n}iz-Rascado, J.~S. Garc{\'\i}a-Sotelo, V.~Weiss, H.~Solano-Lira,
  I.~Mart{\'\i}nez-Flores, A.~Medina-Rivera, {\em et~al.}, ``Regulondb v8. 0:
  omics data sets, evolutionary conservation, regulatory phrases,
  cross-validated gold standards and more,'' {\em Nucleic Acids Research},
  vol.~41, no.~D1, pp.~D203--D213, 2013.

\bibitem{ecoli_thieffry}
D.~Thieffry, A.~M. Huerta, E.~P{\'e}rez-Rueda, and J.~Collado-Vides, ``From
  specific gene regulation to genomic networks: a global analysis of
  transcriptional regulation in escherichia coli,'' {\em Bioessays}, vol.~20,
  no.~5, pp.~433--440, 1998.

\bibitem{UriAlon}
R.~Milo, S.~Itzkovitz, N.~Kashtan, R.~Levitt, S.~Shen-Orr, I.~Ayzenshtat,
  M.~Sheffer, and U.~Alon, ``Superfamilies of evolved and designed networks,''
  {\em Science}, vol.~303, no.~5663, pp.~1538--1542, 2004.

\bibitem{yeast_harbison}
C.~T. Harbison, D.~B. Gordon, T.~I. Lee, N.~J. Rinaldi, K.~D. Macisaac, T.~W.
  Danford, N.~M. Hannett, J.-B. Tagne, D.~B. Reynolds, J.~Yoo, {\em et~al.},
  ``Transcriptional regulatory code of a eukaryotic genome,'' {\em Nature},
  vol.~431, no.~7004, pp.~99--104, 2004.

\bibitem{yeast_costanzo}
M.~C. Costanzo, M.~E. Crawford, J.~E. Hirschman, J.~E. Kranz, P.~Olsen, L.~S.
  Robertson, M.~S. Skrzypek, B.~R. Braun, K.~L. Hopkins, P.~Kondu, {\em
  et~al.}, ``Ypd™, pombepd™ and wormpd™: model organism volumes of the
  bioknowledge™ library, an integrated resource for protein information,''
  {\em Nucleic Acids Research}, vol.~29, no.~1, pp.~75--79, 2001.

\bibitem{paeruginosa_galan}
E.~Gal{\'a}n-V{\'a}squez, B.~Luna, and A.~Mart{\'\i}nez-Antonio, ``The
  regulatory network of pseudomonas aeruginosa,'' {\em Microb Inform Exp},
  vol.~1, no.~1, pp.~3--3, 2011.

\bibitem{mtuberculosis_sanz}
J.~Sanz, J.~Navarro, A.~Arbu{\'e}s, C.~Mart{\'\i}n, P.~C. Mariju{\'a}n, and
  Y.~Moreno, ``The transcriptional regulatory network of mycobacterium
  tuberculosis,'' {\em PloS one}, vol.~6, no.~7, p.~e22178, 2011.

\bibitem{jeong2000large}
H.~Jeong, B.~Tombor, R.~Albert, Z.~N. Oltvai, and A.-L. Barab{\'a}si, ``The
  large-scale organization of metabolic networks,'' {\em Nature}, vol.~407,
  no.~6804, pp.~651--654, 2000.

\bibitem{CElegans_neural}
J.~G. White, E.~Southgate, J.~N. Thompson, and S.~Brenner, ``The structure of
  the nervous system of the nematode caenorhabditis elegans,'' {\em Phil.
  Trans. R. Soc. London}, vol.~314, pp.~1--340, 1986.

\bibitem{WattsStrogatz}
D.~J. Watts and S.~H. Strogatz, ``Collective dynamics of `small-world'
  networks,'' {\em Nature}, vol.~393, pp.~440--442, 1998.

\bibitem{Gnutella_1}
J.~Leskovec, J.~Kleinberg, and C.~Faloutsos, ``Graph evolution: {D}ensification
  and shrinking diameters,'' {\em ACM Transactions on Knowledge Discovery from
  Data}, vol.~1, 2007.

\bibitem{Gnutella_2}
M.~Ripeanu, I.~Foster, and A.~Iamnitchi, ``Mapping the gnutella network:
  {P}roperties of large-scale peer-to-peer systems and implications for system
  design,'' {\em IEEE Internet Computing Journal}, 2002.

\bibitem{Pajek_trade}
W.~de~Nooy, A.~Mrvar, and V.~Batagelj, {\em Exploratory Social Network Analysis
  with {P}ajek}.
\newblock Cambridge: Cambridge University Press, 2004.

\bibitem{DrSeuss}
{Dr Seuss}, {\em Green Eggs and Ham}.
\newblock Random House, 1960.

\end{thebibliography}
%
%

\newpage

\section*{Supplementary Material}

\subsection*{Network data}


The main text makes use of the same set of $62$ empirical networks analysed in Ref. \cite{Johnson_looplessness}.
These include food webs, gene regulatory networks, metabolic networks, a neural network, trade networks, a P2P file sharing network, and 
a network of word adjacencies. All data are available online at:\\ 
\url{https://www.samuel-johnson.org/data}.\\
One can also find on this website the C++ code used for all analyses and simulations performed for the main article.

The tables below list a series of properties for each network, along with references to the original data sources. The captions also include
links to other websites where the data can also be found.

\renewcommand{\tablename}{Table S} 
\setcounter{table}{0}

\begin{center}
 \begin{longtable}{@{}llccccccccc}

Food	web	&	$N$	&	$B$	&	$\langle k\rangle$	&	$q$	&	$q/\tilde{q}$	&	$\tau$	&	$\rho$	&	$\Phi$	&	$d_F$	&	Ref.\\		
\hline																							
\hline																							
Benguela	Current	&	29	&	2	&	6.76	&	0.69	&	0.15	&	0.5	&	2	&	0.1	&	0.97	&	\cite{benguela}	\\	
Berwick	Stream	&	77	&	35	&	3.12	&	0.18	&	0.53	&	-12.21	&	0	&	0	&	1	&	\cite	{streams5,streams6,streams7}	\\
Blackrock	Stream	&	86	&	49	&	4.36	&	0.19	&	0.57	&	-9.51	&	0	&	0	&	1	&	\cite	{streams5,streams6,streams7}	\\
Bridge Brook	Lake	&	25	&	8	&	5.08	&	0.53	&	0.36	&	-0.53	&	1	&	0.08	&	0.99	&	\cite{bridge}	\\	
Broad	Stream	&	94	&	53	&	6	&	0.14	&	0.49	&	-20.1	&	0	&	0	&	1	&	\cite	{streams5,streams6,streams7}	\\
Canton	Creek	&	102	&	54	&	6.82	&	0.15	&	0.57	&	-14.52	&	0	&	0	&	1	&	\cite{canton}	\\	
Caribbean	Reef	&	50	&	3	&	10.7	&	0.94	&	0.33	&	1.73	&	7.8	&	0.56	&	0.9	&	\cite{reef}	\\	
Cayman	Islands	&	242	&	10	&	15.55	&	0.77	&	0.24	&	1.22	&	0	&	0	&	1	&	\cite{caribbean_2005}	\\	
Catlins	Stream	&	48	&	14	&	2.29	&	0.2	&	0.41	&	-10.9	&	0	&	0	&	1	&	\cite	{streams5,streams6,streams7}	\\
Chesapeake	Bay	&	31	&	5	&	2.16	&	0.45	&	0.33	&	-1.81	&	0	&	0	&	1	&	\cite{chesapeake1,chesapeake2}	\\	
Coachella	Valley	&	29	&	3	&	8.38	&	1.2	&	0.48	&	1.63	&	5.48	&	0.38	&	0.89	&	\cite{coachella}	\\	
Coweeta	1	&	58	&	28	&	2.17	&	0.3	&	0.64	&	-3.39	&	0	&	0	&	1	&	\cite	{streams5,streams6,streams7}	\\
Coweeta	17	&	71	&	38	&	2.08	&	0.24	&	0.6	&	-5.94	&	0	&	0	&	1	&	\cite	{streams5,streams6,streams7}	\\
Dempsters	(Au)	&	83	&	46	&	4.99	&	0.21	&	0.57	&	-7.42	&	0	&	0	&	1	&	\cite	{streams5,streams6,streams7}	\\
Dempsters	(Sp)	&	93	&	50	&	5.78	&	0.13	&	0.38	&	-27.07	&	0	&	0	&	1	&	\cite	{streams5,streams6,streams7}	\\
Dempsters	(Su)	&	107	&	50	&	9.02	&	0.27	&	0.57	&	-3.51	&	0.01	&	0	&	1	&	\cite	{streams5,streams6,streams7}	\\
El Verde	Rainforest	&	155	&	28	&	9.72	&	1.01	&	0.45	&	2.09	&	10.12	&	0.45	&	0.94	&	\cite{el_verde}		\\
German	Stream	&	84	&	48	&	4.19	&	0.2	&	0.47	&	-9.35	&	0	&	0	&	1	&	\cite	{streams5,streams6,streams7}	\\
Healy	Stream	&	96	&	47	&	6.6	&	0.22	&	0.53	&	-6.34	&	0	&	0	&	1	&	\cite	{streams5,streams6,streams7}	\\
Kyeburn	Stream	&	98	&	58	&	6.42	&	0.18	&	0.62	&	-9.39	&	0	&	0	&	1	&	\cite	{streams5,streams6,streams7}	\\
LilKyeburn	Stream	&	78	&	42	&	4.81	&	0.23	&	0.53	&	-5.97	&	0	&	0	&	1	&	\cite	{streams5,streams6,streams7}	\\
Little Rock	Lake	&	92	&	12	&	10.7	&	0.67	&	0.22	&	1.06	&	5.66	&	0.23	&	0.97	&		\cite{little_rock}	\\
Lough	Hyne	&	349	&	49	&	14.62	&	0.6	&	0.37	&	0.85	&	2.56	&	0.03	&	1	&	\cite{lough_hyne_1,lough_hyne_2}	\\	
Martins	Stream	&	105	&	48	&	3.27	&	0.32	&	0.58	&	-2.56	&	0	&	0	&	1	&	\cite	{streams5,streams6,streams7}	\\
Narrowdale	Stream	&	71	&	28	&	2.17	&	0.23	&	0.5	&	-7.45	&	0	&	0	&	1	&	\cite	{streams5,streams6,streams7}	\\
NE	Shelf	&	79	&	2	&	17.44	&	0.73	&	0.13	&	1.57	&	4.32	&	0.34	&	0.98	&	\cite{shelf}	\\	
North Col	Stream	&	78	&	25	&	3.09	&	0.28	&	0.52	&	-4.52	&	0	&	0	&	1	&	\cite	{streams5,streams6,streams7}	\\
Powder	Stream	&	78	&	32	&	3.44	&	0.22	&	0.47	&	-8.32	&	0	&	0	&	1	&	\cite	{streams5,streams6,streams7}	\\
Scotch	Broom	&	85	&	1	&	2.58	&	0.4	&	0.14	&	-2.08	&	0	&	0	&	1	&	\cite{broom}	\\	
Skipwith	Pond	&	25	&	1	&	7.56	&	0.61	&	0.15	&	0.2	&	2	&	0.12	&	0.98	&	\cite{skipwith}	\\	
St Marks	Estuary	&	48	&	6	&	4.54	&	0.63	&	0.37	&	0.26	&	0	&	0	&	1	&	\cite{st_marks}	\\	
St Martin	Island	&	42	&	6	&	4.88	&	0.59	&	0.32	&	-0.05	&	0.01	&	0	&	1	&	\cite{st_martin}	\\	
Stony	Stream	&	109	&	61	&	7.59	&	0.15	&	0.55	&	-14.66	&	0	&	0	&	1	&	\cite{stony}	\\	
Stony	Stream 2	&	112	&	63	&	7.41	&	0.15	&	0.55	&	-14.72	&	0	&	0	&	1	&	\cite	{streams5,streams6,streams7}	\\
Sutton	(Au)	&	80	&	49	&	4.19	&	0.15	&	0.66	&	-13.27	&	0	&	0	&	1	&	\cite	{streams5,streams6,streams7}	\\
Sutton	(Sp)	&	74	&	50	&	5.28	&	0.1	&	0.56	&	-35.01	&	0	&	0	&	1	&	\cite	{streams5,streams6,streams7}	\\
Sutton	(Su)	&	87	&	63	&	4.87	&	0.28	&	0.89	&	-1.59	&	0	&	0	&	1	&	\cite	{streams5,streams6,streams7}	\\
Troy	Stream	&	77	&	40	&	2.35	&	0.19	&	0.37	&	-12.16	&	0	&	0	&	1	&	\cite	{streams5,streams6,streams7}	\\
UK	Grassland	&	61	&	8	&	1.59	&	0.4	&	0.18	&	-3.03	&	0	&	0	&	1	&	\cite{grass}	\\	
Venlaw	Stream	&	66	&	30	&	2.83	&	0.23	&	0.54	&	-6.72	&	0	&	0	&	1	&	\cite	{streams5,streams6,streams7}	\\
Weddel	Sea	&	483	&	61	&	31.71	&	0.72	&	0.55	&	2.63	&	22.91	&	0.2	&	0.98	&	\cite{weddell_sea}	\\	
Ythan	Estuary	&	82	&	5	&	4.77	&	0.42	&	0.15	&	-1.32	&	1	&	0.02	&	1	&	\cite{Ythan96}	\\

\caption{
 Details of 42 food webs used in the main text.
 Columns are for number of nodes $N$, number of basal nodes $B$,
 mean degree $\langle k\rangle$,
incoherence parameter $q$, ratio of $q$ to its basal-ensemble expectation $\tilde{q}$, 
loop exponent $\tau$, spectral radius $\rho$, proportion of non-basal vertices in the largest strongly connected component $\Phi$,
normalised deviation from normality $d_F$, and references to the data sources.
Many of the data are available online at:
\url{https://www.nceas.ucsb.edu/interactionweb/html/thomps_towns.html} 
}
\label{table_foodwebs}
 \end{longtable}
\end{center}

\begin{center}
 \begin{longtable}{@{}llccccccccc}

Gene	regulatory	network	&	$N$	&	$B$	&	$\langle k\rangle$	&	$q$	&	$q/\tilde{q}$	&	$\tau$	&	$\rho$	&	$\Phi$	&	$d_F$	&	Ref.\\	
\hline																							
\hline																							
Human	(healthy)		&	4071	&	4004	&	2.08	&	0.08	&	0.99	&	-1.54	&	1	&	0	&	1	&	\cite{human_gerstein,BQS}	\\
Human	(cancer)		&	4049	&	3967	&	2.89	&	0.08	&	1	&	-0.16	&	2.54	&	0	&	1	&	\cite{human_gerstein,BQS}	\\
{\it	E.	coli} (Salgado)	&	1470	&	1316	&	1.98	&	0.23	&	1.03	&	0.65	&	1.62	&	0	&	1	&	\cite{ecoli_salgado,BQS}	\\
{\it	E.	coli} (Thieffry)	&	418	&	312	&	1.24	&	0.27	&	0.88	&	-2.54	&	0	&	0	&	1	&	\cite{ecoli_thieffry,UriAlon}	\\
{\it	S.	cerevisiae} (Harbison)	&	2933	&	2764	&	2.1	&	0.17	&	0.98	&	-0.38	&	1	&	0	&	1	&	\cite{yeast_harbison,BQS}	\\
{\it	S.	cerevisiae} (Costanzo)	&	688	&	557	&	1.57	&	0.25	&	1.04	&	-0.31	&	1.32	&	0	&	1	&	\cite{yeast_costanzo,UriAlon}	\\
{\it	P.	aeruginosa}	&	691	&	606	&	1.43	&	0.3	&	1	&	0.58	&	1.41	&	0	&	0.99	&	\cite{paeruginosa_galan,BQS}	\\
{\it	M.	tuberculosis}	&	1624	&	1542	&	1.95	&	0.17	&	1.02	&	0.99	&	2	&	0	&	1	&	\cite{mtuberculosis_sanz,BQS}	\\

\caption{
Details of eight gene regulatory networks (GRN) used in the main text.
The E. coli (Salgado) and Yeast (Harbison) are available online at: \url{http://wws.weizmann.ac.il/mcb/UriAlon/download/collection-complex-networks}.
The others were shared with us by Luca Albergante, and some of them can be obtained from various websites: \url{http://regulondb.ccg.unam.mx/} (E. coli, Salgado);
\url{http://younglab.wi.mit.edu/regulatory_code} (Yeast, Harbison);
\url{http://www.genome.gov/ENCODE/} (Human, both the non-cancer GM12878 cell line and the K562 leukaemia cell line).
Columns as in Table S\ref{table_foodwebs}.
}
\label{table_gene_nets}
 \end{longtable}
\end{center}


\begin{center}
 \begin{longtable}{@{}llccccccccc}

Metabolic	network		&	$N$	&	$B$	&	$\langle k\rangle$	&	$q$	&	$q/\tilde{q}$	&	$\tau$	&	$\rho$	&	$\Phi$	&	$d_F$	&		Ref.\\	
\hline																								
\hline																								
{\it	A.	fulgidus}	&	1267	&	36	&	2.38	&	13.79	&	1.88	&	2.35	&	7.62	&	0.92	&	0.63	&		\cite{jeong2000large}	\\
{\it	M.	thermoautotrophicum}	&	1111	&	30	&	2.43	&	12.17	&	1.77	&	2.31	&	7.59	&	0.92	&	0.64	&		\cite{jeong2000large}	\\
{\it	M.	jannaschii}	&	1081	&	32	&	2.4	&	12.47	&	1.86	&	2.27	&	7.53	&	0.92	&	0.63	&		\cite{jeong2000large}	\\
{\it	C.	pneumoniae}	&	386	&	20	&	2.05	&	8.98	&	1.62	&	1.69	&	5.57	&	0.74	&	0.63	&		\cite{jeong2000large}	\\
{\it	C.	trachomatis}	&	446	&	19	&	2.11	&	11.77	&	1.95	&	1.79	&	6.07	&	0.78	&	0.63	&		\cite{jeong2000large}	\\
{\it	S.	cerevisiae} (yeast)	&	1510	&	43	&	2.54	&	14.61	&	1.73	&	2.66	&	9.15	&	0.91	&	0.64	&		\cite{jeong2000large}	\\
{\it	C.	elegans}	&	1172	&	40	&	2.44	&	13.29	&	1.86	&	2.44	&	8	&	0.88	&	0.63	&		\cite{jeong2000large}	\\

\caption{
Details of seven metabolic networks used in the main text, downloaded from 
\url{http://www3.nd.edu/~networks/resources.htm}.
Columns as in Table S\ref{table_foodwebs}.
}
\label{table_meta_nets}
 \end{longtable}
\end{center}

\begin{center}
 \begin{longtable}{@{}llccccccccc}
Network (miscellaneous) 	&	$N$	&	$B$	&	$\langle k\rangle$	&	$q$	&	$q/\tilde{q}$	&	$\tau$	&	$\rho$	&	$\Phi$	&	$d_F$	&	Ref.	\\
\hline
\hline
Neural ({\it C.elegans})	&	297	&	3	&	7.9	&	1.49	&	0.42	&	2.17	&	9.15	&	0.78	&	0.84	&	\cite{CElegans_neural,WattsStrogatz}	\\
P2P (Gnutella 2008)	&	6301	&	3836	&	3.3	&	0.98	&	0.98	&	1.49	&	5.12	&	0.33	&	0.94	&	\cite{Gnutella_1,Gnutella_2}	\\
Trade (manuf. goods)	&	24	&	2	&	12.92	&	4.24	&	1.14	&	2.68	&	14.3	&	0.92	&	0.48	&	\cite{Pajek_trade}	\\
Trade (minerals)	&	24	&	3	&	5.63	&	4.04	&	1.02	&	2.05	&	7.38	&	0.88	&	0.61	&	\cite{Pajek_trade}	\\
Words	&	50	&	16	&	2.02	&	2.04	&	1.01	&	1.31	&	3.17	&	0.52	&	0.82	&	\cite{DrSeuss,Johnson_looplessness}	\\

\caption{
Details of five other networks used in the main text.
The network of words was obtained from 
{\it Green Eggs and Ham} \cite{DrSeuss}, 
as described in Ref. \cite{Johnson_looplessness},
and is available at \url{https://www.samuel-johnson.org/data}.
The other data can be found on various websites:
\url{http://www-personal.umich.edu/~mejn/netdata/} (neural network);
\url{https://snap.stanford.edu/data/p2p-Gnutella08.html} (P2P network); and
\url{http://vlado.fmf.uni-lj.si/pub/networks/data/esna/metalWT.htm} (trade networks).
Columns as in Table S\ref{table_foodwebs}
}
\label{table_misc_nets}
 \end{longtable}
\end{center}

\end{document}